\documentclass[english,pra,showpacs,twocolumn]{revtex4-1}
\usepackage[T1]{fontenc}
\usepackage[latin9]{inputenc}
\usepackage{xcolor}
\usepackage{amsmath}
\usepackage{graphicx}

\makeatletter
\usepackage[colorlinks = true,
            linkcolor =blue,
            urlcolor  =blue,
            citecolor = blue,
            anchorcolor = blue]{hyperref}

\makeatother

\usepackage{babel}
\begin{document}
\title{Studying the squeezing effect and phase space distribution of single-
photon- added coherent state using postselected von Neumann measurement}
\author{Wen Jun Xu, Taximaiti Yusufu}
\author{Yusuf Turek}
\email{yusufu1984@hotmail.com}

\affiliation{School of Physics and Electronic Engineering, Xinjiang Normal University,
Urumqi, Xinjiang 830054, China}
\date{\today}
\begin{abstract}
In this paper, ordinary and amplitude-squared squeezing as well as
Wigner functions of single-photon-added coherent state after postselected
von Neumann measurements are investigated. The analytical results
show that the von Neumann type measurement which is characterized
by post-selection and weak value can significantly change the squeezing
feature of single-photon-added coherent state. It is also found that
the postselected measurement can increase the nonclassicality of the
original state in strong measurement regimes. It is anticipated that
this work could may provide an alternate and effective methods to
solve state optimization problems based on the postselected von Neumann
measurement technique.
\end{abstract}
\pacs{42.50.-p, 03.65.-w, 03.65.Ta}
\maketitle

\section{\label{sec:1}Introduction}

States which possess nonclassical features are an important resources
for quantum information processing and the investigation of fundamental
problems in quantum theory. It has been shown that squeezed states
of radiation fields has been can be considered truly quantum \citep{2}.
In recent years studies concerning squeezing especially quadrature
squeezing of radiation fields has seen considerable attention as it
may have applcaition in optical communication and information theory
\citep{10,11,12,13,14,15,16,17,18,7,8,9}, gravitatioanl wave detection
\citep{19}, quantum teleportation \citep{19,20,21,22,23,24,25,26,27},
dense coding \citep{28}, resonance fluorscence \citep{29}, and quantum
cryptography \citep{30}. Furthermore, with the rapid development
of the techniques for making higher-order correlation measurements
in quantum optics and laser physics, the high-order squeezing effects
of radiation fields have also became a hot topic in state optimization
researches. Higher-order squeezing of radiation fields was first introduced
by Hong and Manel \citep{31} in 1985, and Hilley \citep{32,33} defined
another type higher-order squeezing, named amplitude- squared squeezing
(ASS) of the electromagnetic field in 1987.\textcolor{brown}{{} }Following
this work the highe-squeezing of radiation fields has been investigated
across many fields of research. \citep{34,35,36,37,38,39,40,41,42,43,44,45,46,47,48,49,50}. 

Squeezing is an inherent feature of nonclassical states, and its improvement
requires optimization. Some states do not initially possess squeezing,
but after undergoing an optimization process, they may possess a pronounced
squeezing effect, The single-photon-added coherent state (SPACS) is
a typical example. SPACAS are created by adding the creation operator
$a^{\dagger}$ to the coherent state, and this optimization changes
the coherent state from semi-classical to a new quantum state which
possess squeezing. Since this state has wide application across many
quantum information processes including quantum communication \citep{RN1923},
quantum key distribution \citep{WANG20171393,RN1920,RN1924,RN1925},
and quantum digital signature \citep{RN1921}, the optimization for
this state is worthy of study, in particular, it may provide new methods
to the implementations related processes. On the other hand, the weak
signal amplification technique proposed in 1988\textcolor{brown}{{}
\citep{59}} by Aharonov, Albert, and Vaidman is widely used in state
optimization and precision measurement problems \citep{75,76,77,78,79,80,81}.
Most recently, one of the authors of this paper investigated the effects
of postselected von Neumann measurement on the properties of single-mode
radiation fields \citep{80,81} and found that postselected von Neumann
measurement changed the photon statistics and quadrature squeezing
of radiation fields for different anomalous weak values and coupling
strengths. However, to the best of our knowledge, the effects of postselected
von Neumann measurement on higher-order squeezing and phase-space
distribution of SPACS have not been previously investigated.

In this work, motivated by our prior work \citep{78,80,81}, we study
the squeezing and Wigner function of SPACS after postselected von
Neumann measurement. In this work, we take the spatial and polarization
degrees of freedom of SPACS as a measuring device (pointer) and system,
respectively, and consider all orders of the time evolution operator.
Following determination of the final state of the pointer, we check
the criteria for existence of squeezing of SPACS, and found that the
postselected measurement has positve effects on squeezing of SPACS
in the weak measurement regime. Furthermore, we investigate the state-distance
and the Wigner function of the SPACS after measurement. We found that
with increasing coupling strength, the original SPACS spoiled significantly,
and the state exhibited more pronounced negative areas as well as
interference structures in phase space after postselected measurement.\textcolor{brown}{{}
}We observed that the postselected von Neumann measurement has positive
effects on its nonclassicality including squeezing effects especially
in the weak measurement regime. These results can be considered a
result of weak value amplification of the weak measurement technique. 

This paper is organized as follows. In Sec. \ref{sec:2}, we introduce
the main concepts of our scheme and derive the final pointer state
after postselected measurement which will be used throughout the study.
In Sec. \ref{sec:3}, we give the details of ordinary squeezing and
ASS effects of the final pointer state. In Sec. \ref{sec:4}, we investigate
the state distance and the Wigner function SPACS after measurement.
A conclusion is given in Sec. \ref{sec:5}. 

\section{\label{sec:2}Model and theory }

In this section, we introduce the basic concepts of postselected von
Neumann measurement and give the expression of the final pointer state
which we use in this paper. We know that every measurement problems
consists of three main parts including a pointer(measuring device),
measuring system and the environment. In the current work, we take
the spatial and polarization degrees of freedom of SPACS as the pointer
and system, respectively. In general, in measurement problems we want
to determine the system information of interest by comparing the state-shifts
of the pointer after measurement finishes, and we do not consider
spoiling of the pointer in the entire measurement process. Here, contrary
to the standard goal of the measurement, we investigate the effects
of pre- and post-selected measurement taken on a beam's polarization(measured
system) on the inherent properties of a beam's spatial component (pointer).
In the measurement process, the system and pointer Hamiltonians doesn't
effect the final read outs, so it is sufficient to only consider their
interaction Hamiltonian for our purposes. According to standard von
Neumann measurement theory, the interaction Hamiltonian between the
system and the pointer takes the form \citep{Edwards1955} 

\begin{equation}
\hat{H}=g(t)\hat{A}\otimes\hat{P}.\label{eq:2}
\end{equation}
Here, $\hat{A}$ is the system observable we want to measure, and
$\hat{P}$ is the momentum operator of the pointer conjugated with
the position operator, $\left[\hat{X},\hat{P}\right]=i$. $g(t)$
is the coupling strength function between the system and pointer and
it is assumed exponentially small except during a period of interaction
time of order $T$, and is normalized according to $\int_{-\infty}^{+\infty}g(t)dt=\int_{0}^{T}g(t)dt=g_{0}$.
In this work, we assume that the system observable $A$ is Pauli $x$
matrix, i.e., 
\begin{equation}
\hat{A}=\hat{\sigma}_{x}=\vert H\rangle\langle V\vert+\vert V\rangle\langle H\vert=\left(\begin{array}{cc}
0 & \ \ 1\\
1 & 0
\end{array}\right)\label{eq:1}
\end{equation}
Here, $\vert H\rangle\equiv(1,0)^{T}$ and $\vert V\rangle\equiv(0,1)^{T}$represent
the horizontal and vertical polarization of the beam, respectively.
We also assume that in our scheme the pointer and measurement system
are initially prepared to 
\begin{equation}
\vert\phi\rangle=\gamma a^{\dagger}\vert\alpha\rangle,\ \ \ \ \gamma=\frac{1}{\sqrt{1+\vert\alpha\vert^{2}}}\label{eq:3-2}
\end{equation}
and 
\begin{equation}
\vert\psi_{i}\rangle=\cos\frac{\varphi}{2}\vert H\rangle+e^{i\delta}\sin\frac{\varphi}{2}\vert V\rangle,\label{eq:3-1-1}
\end{equation}
respectively. Here, $\alpha=re^{i\theta}$ and $\delta\in[0,2\pi]$
and $\varphi\in[0,\pi)$. Here, we are reminded that in weak measurement
theory, the interaction strength between the system and measurement
is weak, and it is enough to only consider the evolution of the unitary
operator up to its first order. However, if we want to connect the
weak and strong measurement and investigate the measurement feedback
of postselected weak measurement, and analyze experimental results
obtained in non-ideal measurements, the full-order evolution of the
unitary operator is needed \citep{72,73,74}, We call this kind of
measurement a postselected von Neumann measurement. Thus, the evolution
operator of this total system corresponding to the interaction Hamiltonian,
Eq. (\ref{eq:2}), is evaluated as 
\begin{equation}
e^{-ig_{0}\sigma_{x}\otimes P}=\frac{1}{2}\left(\hat{I}+\hat{\sigma}_{x}\right)\otimes D\left(\frac{s}{2}\right)+\frac{1}{2}\left(\hat{I}-\hat{\sigma}_{x}\right)\otimes D\left(-\frac{s}{2}\right)\label{eq:7}
\end{equation}
since$\hat{\sigma}_{x}^{2}=1$. Here,$s=\frac{g_{0}}{\sigma}$ is
the ratio between the coupling strength and beam width, and it can
characterize the measurement types i.e. the measurement is considered
a weak measurement (strong measurement) if $s<1$ ($s>1$). $D(\frac{s}{2})$
is the displacement operator defined as $D(\alpha)=e^{\alpha\hat{a}^{\dagger}-\alpha^{\ast}\hat{a}}$.
The results of our current research are valid for weak and strong
measurement regimes since we take into account the all orders of the
time evolution operator, Eq. (\ref{eq:7}). In the above calculation
we use the definition of the momentum operator represented in Fock
space in terms of an annihilation (creation) operator $\hat{a}$ ($\hat{a}^{\dagger}$),
i.g., 
\begin{equation}
\hat{P}=\frac{i}{2\sigma}\left(a^{\dagger}-a\right)\label{eq:P}
\end{equation}
where $\sigma$ is the width of the beam. Thus, the total state of
the system, $\vert\psi_{i}\rangle\otimes\vert\phi\rangle$, after
the time evolution becomes 
\begin{align}
\vert\Psi\rangle & =e^{-ig_{0}\sigma_{x}\otimes P}\vert\psi_{i}\rangle\otimes\vert\phi\rangle\nonumber \\
 & =\frac{1}{2}\left[\left(\hat{I}+\hat{\sigma}_{x}\right)\otimes D\left(\frac{s}{2}\right)+\left(\hat{I}-\hat{\sigma}_{x}\!\right)\otimes D\left(\!\frac{-s}{2}\right)\right]\!\vert\psi_{i}\rangle\otimes\vert\phi\rangle\label{eq:8}
\end{align}
After we take a strong projective measurement of the polarization
degree of the beam with posts-elected state $\vert\psi_{f}\rangle=\vert H\rangle$,
the above total system state gives us the final state of the pointer,
and its normalized expression reads as 
\begin{equation}
\vert\Phi\rangle=\frac{\kappa}{\sqrt{2}}\left[\left(1+\langle\sigma_{x}\rangle_{w}\right)D\left(\frac{s}{2}\right)+\left(1-\langle\sigma_{x}\rangle_{w}\right)D\left(-\frac{s}{2}\right)\right]\vert\phi\rangle.\label{eq:10-2}
\end{equation}
 Here, 
\begin{align}
\kappa^{-2} & =\!1+\!\vert\langle\sigma_{x}\rangle\vert^{2}\!+\gamma^{2}e^{-\frac{s^{2}}{2}}Re[(1+\langle\sigma_{x}\rangle_{w}^{\ast})(1-\!\langle\sigma_{x}\rangle_{w})\times\nonumber \\
 & (\gamma^{-2}-\text{\ensuremath{s^{2}+\alpha s-\alpha^{\ast}s}})e^{2si\Im(\alpha)}]\label{eq:11-1}
\end{align}
 is the normalization coefficient, and the weak value of the system
observable $\hat{\sigma}_{x}$ is given by 
\begin{equation}
\langle\sigma_{x}\rangle_{w}=\frac{\langle\psi_{f}\vert\sigma_{x}\vert\psi_{i}\rangle}{\langle\psi_{f}\vert\psi_{i}\rangle}=e^{i\delta}\tan\frac{\varphi}{2}.\label{eq:3}
\end{equation}
 In general, the expectation value of $\sigma_{x}$ is bounded $-1\le\langle\sigma_{x}\rangle\le1$
for any associated system state. However, as we see in Eq. (\ref{eq:3}),
the weak values of the observable $\sigma_{x}$ can take arbitrary
large numbers with small successful post-selection probability $P_{s}=\vert\langle\psi_{f}\vert\psi_{i}\rangle\vert^{2}=\cos^{2}\frac{\varphi}{2}$.
This weak value feature is used to amplify very weak but useful information
on various of related physical systems. 

The state given in Eq. ( \ref{eq:10-2}) is a spoiled version of SPACS
after postselected measurement. In the next sections, we study squeezing
effects, and nonclassicality features characterized by the Wigner
function. 

\section{\label{sec:3}Ordinary and amplitude square squeezing}

In this section, we check the ordinary (first-order) and ASS (second
order) squeezing effects of SPACS after postselected von Neumann measurement.The
squeezing effect is one of the non-classical phenomena unique to the
quantum light field. The squeezing reflects the non-classical statistical
properties of the optical field by a noise component lower than that
of the coherent state. In other words, the noise of an orthogonal
component of the squeezed light is lower than the noise of the corresponding
component of the coherent state light field. In practice, if this
component is used to transmit information, a higher signal-to-noise
ratio can be obtained than that of the coherent state. Consider a
single mode of electromagnetic field of frequency $\text{\ensuremath{\omega}}$
with creation and annihilation operator $a^{\dagger},$$a$. The quadrature
and square of the field mode amplitude can be defined by operators
$X_{\theta}$ and $Y_{\theta}$ as \citep{Agarwal2013} 
\begin{equation}
X_{\theta}\equiv\frac{1}{2}\left(ae^{-i\theta}+a^{\dagger}e^{i\theta}\right)\label{eq:11}
\end{equation}
and 
\begin{equation}
Y_{\theta}\equiv\frac{1}{2}\left(a^{2}e^{-i\theta}+a^{\dagger2}e^{i\theta}\right),\label{eq:12}
\end{equation}
respectively. For these operators, if $\triangle X_{\theta}\equiv X_{\theta}-\langle X_{\theta}\rangle$,
$\triangle Y_{\theta}\equiv Y_{\theta}-\langle Y_{\theta}\rangle$,
the minimum variances are \citep{2013} 
\begin{equation}
\langle(\triangle X_{\theta})^{2}\rangle_{min}=\frac{1}{4}+\frac{1}{2}\left[\left(\langle a^{\dagger}a\rangle-\vert\langle a\rangle\vert^{2}\right)-\vert\langle a^{2}\rangle-\langle a\rangle^{2}\vert\right]\label{eq:13}
\end{equation}
 
\begin{align}
\langle(\triangle Y_{\theta})^{2}\rangle_{min} & =\langle a^{\dagger}a+\frac{1}{2}\rangle\label{eq:14}\\
 & +\frac{1}{2}\left[\langle a^{\dagger2}a^{2}\rangle-\vert\langle a^{2}\rangle\vert^{2}-\vert\langle a^{4}\rangle-\langle a^{2}\rangle^{2}\vert\right]\nonumber 
\end{align}
 where $a$ and $a^{\dagger}$ are annihilation and creation operators
of the radiation field. If $\langle(\triangle X_{\theta})^{2}\rangle_{min}<\frac{1}{4}$
, $X_{\theta}$ is said to be ordinary squeezed and if $\langle(\triangle Y_{\theta})^{2}\rangle_{min}<\langle a^{\dagger}a+\frac{1}{2}\rangle$,
$Y_{\theta}$ is said to be ASS. These conditions can be rewritten
as 
\begin{equation}
S_{os}=\langle a^{\dagger}a\rangle-\vert\langle a\rangle\vert^{2}-\vert\langle a^{2}\rangle-\langle a\rangle^{2}\vert<0\label{eq:15}
\end{equation}
 
\begin{equation}
S_{ass}=\langle a^{\dagger2}a^{2}\rangle-\vert\langle a^{2}\rangle\vert^{2}-\vert\langle a^{4}\rangle-\langle a^{2}\rangle^{2}\vert<0.\label{eq:16}
\end{equation}
 Thus, the system characterized by any wave function may exhibit non-classical
features if it satisfies Eqs. (\ref{eq:15}-\ref{eq:16}). To achieve
our goal, we first have to calculate the above related quantities
and their explicit expressions under the state $\vert\Phi\rangle$.these
are listed below. 

\begin{widetext}

1.The expectation value $\langle a^{\dagger}a\rangle$ under the state
$\vert\Psi\rangle$ is given by

\begin{align}
\langle a^{\dagger}a\rangle & =\vert\kappa\vert^{2}\left\{ \vert1+\langle\sigma_{x}\rangle_{w}\vert^{2}t_{1}(s)+\vert1-\langle\sigma_{x}\rangle_{w}\vert^{2}t_{1}(-s)+2Re[\left(1-\langle\sigma_{x}\rangle_{w}\right)\left(1+\langle\sigma_{x}\rangle_{w}\right)^{\ast}t_{3}(s)]\right\} \label{eq:a^a}
\end{align}
where
\[
t_{1}(s)=\gamma^{2}\left(\left(2+\vert\alpha\vert^{4}+s\vert\alpha\vert^{2}\right)Re(\alpha)+3\alpha\alpha^{\ast}+1\right)+\frac{s^{2}}{4}
\]
 and 

\begin{align*}
t_{3}(s) & =\frac{1}{4}\gamma^{2}e^{2isIm(\alpha)}e^{-\frac{s^{2}}{2}}(4\vert\alpha\vert^{4}-6s\alpha\vert\alpha\vert^{2}+2(6\alpha\alpha^{\ast}+s\alpha^{\ast2}(3\alpha+s)\\
 & +sRe(\alpha)(8-9s\alpha-3s^{2}))+11\alpha^{2}s^{2}+s^{4}+6\alpha s^{3}-5s^{2}-16\alpha s+4)
\end{align*}
respectively.

2.The expectation value $\langle a\rangle$ under the state  $\vert\Psi\rangle$
is given by

\begin{align}
\langle a\rangle & =\vert\kappa\vert^{2}\gamma^{2}\{\vert1+\langle\sigma_{x}\rangle_{w}\vert^{2}\left[2\alpha+\alpha\vert\alpha\vert^{2}+\frac{s}{2\text{\ensuremath{\gamma^{2}}}}\right]+\vert1-\langle\sigma_{x}\rangle_{w}\vert^{2}\left[2\alpha+\alpha\vert\alpha\vert^{2}-\frac{s}{2\gamma^{2}}\right]\nonumber \\
 & +\left(1-\langle\sigma_{x}\rangle_{w}\right)\left(1+\langle\sigma_{x}\rangle_{w}\right)^{\ast}w_{1}(s)++\left(1+\langle\sigma_{x}\rangle_{w}\right)\left(1-\langle\sigma_{x}\rangle_{w}\right)^{\ast}w_{1}(-s)]\}\label{eq:a}
\end{align}
where

\[
w_{1}(s)=\frac{1}{2}e^{2isIm(\alpha)}e^{-\frac{s^{2}}{2}}\left(4\alpha+\alpha^{\ast}(s-2\alpha)(s-\alpha)+2\alpha^{2}s+s^{3}-3\alpha s^{2}-3s\right)
\]
 3.The expectation value $\langle a^{2}\rangle$ under the state  $\vert\Psi\rangle$
is given by

\begin{align}
\langle a^{2}\rangle & =\vert\kappa\vert^{2}\{\vert1+\langle\sigma_{x}\rangle_{w}\vert^{2}q_{1}(s)+\vert1-\langle\sigma_{x}\rangle_{w}\vert^{2}q_{1}(-s)+\left(1-\langle\sigma_{x}\rangle_{w}\right)\left(1+\langle\sigma_{x}\rangle_{w}\right)^{\ast}q_{2}(s)\nonumber \\
 & +\left(1+\langle\sigma_{x}\rangle_{w}\right)\left(1-\langle\sigma_{x}\rangle_{w}\right)^{\ast}q_{2}(-s)\}\label{eq:a2}
\end{align}
where

\[
q_{1}(s)=\frac{1}{4}\gamma^{2}(2\alpha+s)(6\alpha+\vert\alpha\vert^{2}(2\alpha+s)+s)
\]
 and 

\begin{align*}
q_{2}(s) & =-\frac{1}{4}e^{2isIm(\alpha)}e^{-\frac{s^{2}}{2}}\gamma^{2}(s-2\alpha)(6\alpha+\alpha^{\ast}(s-2\alpha)(s-\alpha)+2\alpha^{2}s+s^{3}-3\alpha s^{2}-5s)
\end{align*}
 respectively. 

4.The expectation value $\langle a^{\dagger2}a^{2}\rangle$ under
the state  $\vert\Psi\rangle$ is given by

\begin{align}
\langle a^{\dagger2}a^{2}\rangle & =\vert\kappa\vert^{2}\{\vert1+\langle\sigma_{x}\rangle_{w}\vert^{2}f_{1}(s)+\vert1-\langle\sigma_{x}\rangle_{w}\vert^{2}f_{2}(-s)+2Re[\left(1-\langle\sigma_{x}\rangle_{w}\right)\left(1+\langle\sigma_{x}\rangle_{w}\right)^{\ast}f_{3}(s)]\}\label{eq:aa}
\end{align}
where 
\begin{align*}
f_{1}(s) & =\frac{1}{2}\gamma^{2}(2\vert\alpha\vert^{6}+s\vert\alpha\vert^{2}((s^{2}+16)Re(\alpha)+sRe(\alpha^{2}))+2\vert\alpha\vert^{4}(2sRe(\alpha)+s^{2}+5)\\
 & +8\alpha^{\ast}\alpha+6s^{2}\alpha^{\ast}\alpha+(2s^{3}+8s)Re(\alpha)+3s^{2}Re(\alpha^{2}))+\frac{s^{4}}{16}+\gamma^{2}s^{2}
\end{align*}
 and 

\begin{align*}
f_{3}(s) & =-\frac{1}{16}\gamma^{2}(s-2\alpha)\left(2\alpha^{*}+s\right)\\
 & \left(2\left(\alpha^{*}\right)^{2}(s-2\alpha)(s-\alpha)+20\vert\alpha\vert^{2}+3s\alpha^{*}(s-2\alpha)(s-\alpha)+28isIm(\alpha)+s^{2}\left(2\alpha^{2}+s^{2}-3\alpha s-9\right)+16e^{-\frac{1}{2}s(s-4iIm[\alpha])}\right)
\end{align*}
respectively.

5.The expectation value $\langle a^{4}\rangle$ under the state  $\vert\Psi\rangle$
is given by

\begin{align}
\langle a^{4}\rangle & =\vert\kappa\vert^{2}\{\vert1+\langle\sigma_{x}\rangle_{w}\vert^{2}h_{1}(s)+\vert1-\langle\sigma_{x}\rangle_{w}\vert^{2}h_{1}(-s)+(1+\langle\sigma_{x}\rangle_{w})^{\ast}(1-\langle\sigma_{x}\rangle_{w})h_{2}(s)\label{eq:a17}\\
 & +\left(1+\langle\sigma_{x}\rangle_{w}\right)\left(1-\langle\sigma_{x}\rangle_{w}\right)^{\ast}h_{2}(-s)\}
\end{align}
 where 
\begin{align*}
h_{1}(s) & =\frac{1}{16}(8\alpha\gamma^{2}\vert\alpha\vert^{2}(\alpha+s)(2\alpha^{2}+s^{2}+2\alpha s)+s^{4}+8\alpha\gamma^{2}(10\alpha^{3}+2s^{3}+9\alpha s^{2}+16\alpha^{2}s))
\end{align*}
 and 
\begin{align*}
h_{2}(s) & =-\frac{1}{16}\gamma^{2}e^{2isIm(\alpha)}e^{-\frac{s^{2}}{2}}(s-2\alpha)^{3}(10\alpha+\alpha^{\ast}(s-2\alpha)(s-\alpha)+2\alpha^{2}s+s^{3}-3\alpha s^{2}-9s)
\end{align*}
 respectively.

\end{widetext}

\begin{figure}
\includegraphics[width=8cm]{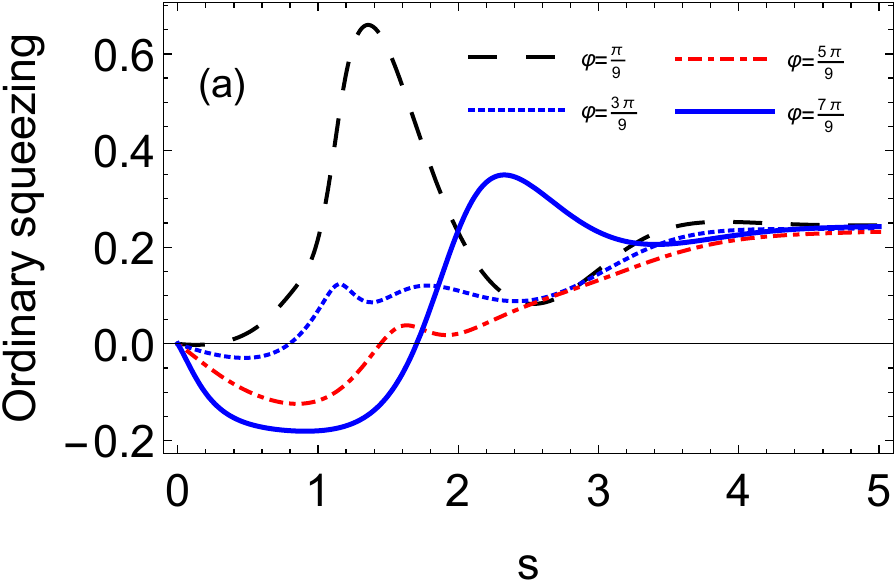}

\includegraphics[width=8cm]{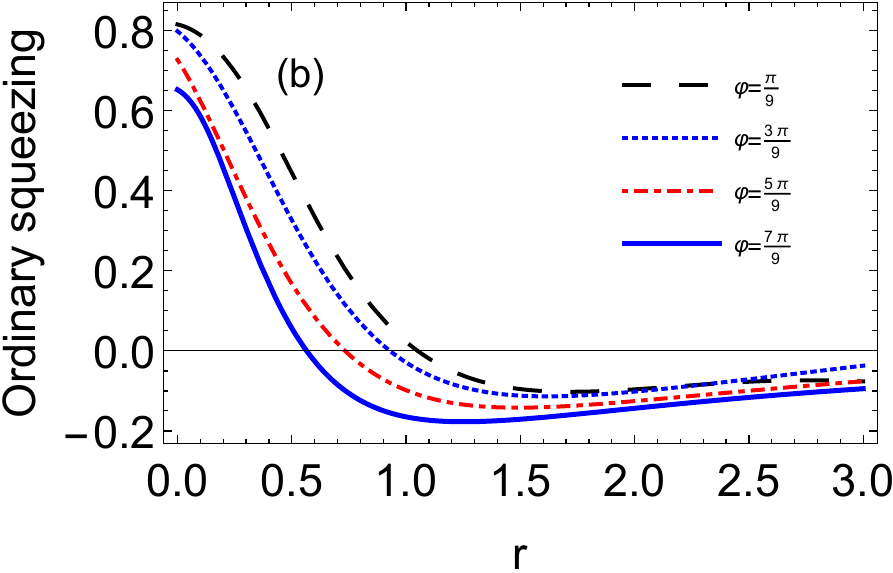}

\caption{\label{Fig1}(Color online) The effects of postselected von Neumann
measurement on ordinary squeezing of SPACS. Fig. \ref{Fig1}(a) shows
the quantity $S_{os}$as a function of coupling strength for different
weak values with fixed coherent state parameter ($r=1$). Fig. \ref{Fig1}(b)
shows quantity $S_{os}$ as a function of coherent state parameter
$r$ for different weak values with fixed coupling strength ($s=0.5$).
Here, we take $\theta=\frac{\pi}{4}$, $\delta=\frac{\pi}{6}$.}
\end{figure}

Using the expression for $S_{os}$, the curves for this quantity are
plotted, and the analytical results are shown in Fig. \ref{Fig1}.
In Fig. \ref{Fig1}(a), we fixed the parameter $r=1$ and plot the
$S_{os}$ as a function of coupling strength $s$ for different weak
values quantified by $\varphi$. As we observed, when there is no
interaction between system and poiner ($s=0$), there is no ordinary
squeezing effect of initial SPACS at the $r=1$ point. However, in
moderate coupling strength regions such as $0<s$<2 , the ordinary
squeezing effect of SPACS is proportional to the weak value, i.e.
the larger the weak value, the better its squeezing effect. From Fig.
\ref{Fig1}(a) we also can see that the ordinary squeezing effect
of the light field gradually disappears and tends to the same value
for different weak values with increasing coupling strength $s$ in
the strong measurement regime. In Fig. \ref{Fig1}(b), we plot $S_{os}$
as a function of the state parameter $r$ in the weak measurement
regime by fixing the coupling strength $s$, i.g, $s=0.5$. It is
very clear from the curves presented in Fig. \ref{Fig1} (b) that
the ordinary squeezing effect of SPACS is increased when increasing
the weak value, especially when $\varphi$ is taken as $\frac{7\pi}{9}$.
Furthermore, along with the increasing $r$ (for $r$ values exceeding
1.5), the squeezing effect of the field for different weak values
tended to be the same. According to von Neumann measurement theory,
when the interaction strength is too large, the system is strongly
measured and the size of the weak value has little impact on the squeezing
effect.This statement can also be observed in Fig. \ref{Fig1}(a)
and (b). In the weak measurement regime the SPACS showed a good ordinary
squeezing effect after postselected measurement with large weak values,
and this can be seen as a result of the signal amplification feature
of the weak measurement technique.

\begin{figure}
\includegraphics[width=8cm]{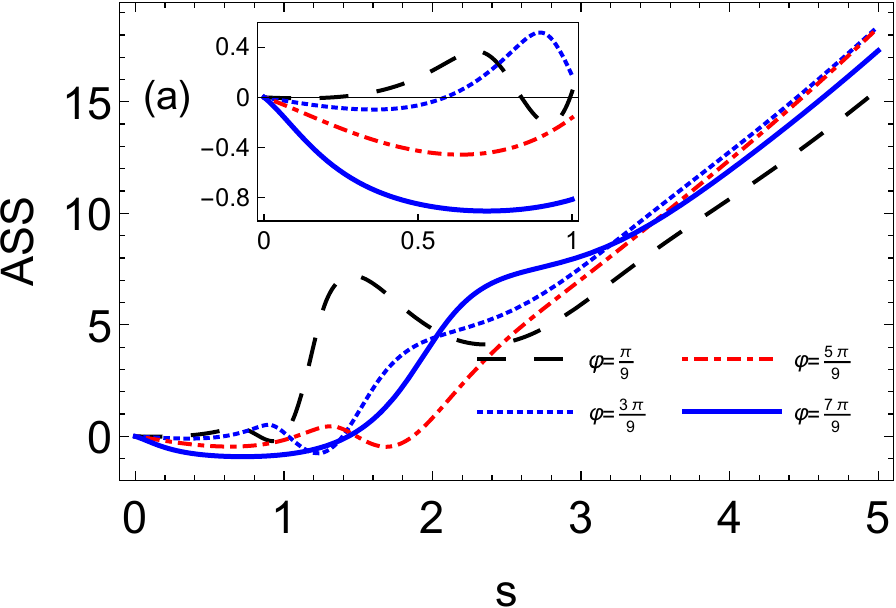}

\includegraphics[width=8cm]{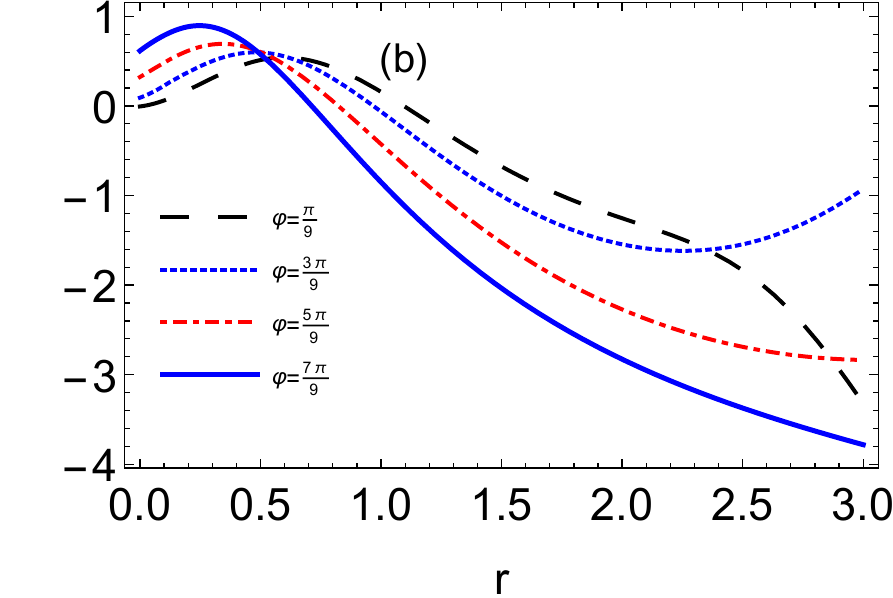}

\caption{\label{Fig2} (Color online) The effects of postselected von Neumann
measurement on ASS of SPACS. (a) the $S_{ass}$ as a function of coupling
strength $s$ for different weak values with fixed coherent state
parameter $r$ ($r=1$); (b) the $S_{ass}$ as a function of coherent
state parameter $r$ for different weak values with fixed weak coupling
strength $s$ ($s=0.5$). Other parameters are the same as those used
in Fig. (\ref{Fig1}).}
\end{figure}

The quantity $S_{ass}$can characterize the ASS of SPACS if it takes
negative values, and in Fig. \ref{Fig2} it is plotted as a function
of various system parameters. As indicated in Fig. \ref{Fig2} (a),
when we fixed the coherent state parameter $r$ ,the $S_{ass}$ can
take negative values in the weak measurement regime ($s<1$) and its
negativity increases when increasing the weak value quantified by
$\varphi$. That is to say, in the weak measurement regime, the magnitude
of the weak value has a linear relationship with the ASS effect of
SPACS, i.g. The larger the weak value, the better the ASS effect.
However, by increasing the coupling strength, the value of $S_{ass}$
became larger than zero and it indicates that there is no ASS effect
on SPACS in the postselected strong measurement regime ($s>1$) no
matter how large the value is taken. In order to further investigate
the ASS of the radiation field in the weak measurement regime, we
plot the $S_{ass}$ as a function of the coherent state parameter
$r$ for different weak values with $s=0.5$. the analytical results
are shown in Fig. \ref{Fig2}b. We can see that when $r$ is relatively
small, there is an ASS effect no matter how large the weak value becomes.
By increasing the system parameter $r$, $S_{ass}$ takes negative
values and its negativity is proportional to $r$. From Fig. \ref{Fig2}a
we can also observe that in the weak measurement regime, the weak
values have positive effects on the ASS of SPACS, and it can also
be considered a result of the weak signal amplification feature of
the postselected weak measurement technique. 

\section{\label{sec:4}State distance and Wigner function }

\begin{figure}
\includegraphics[width=8cm]{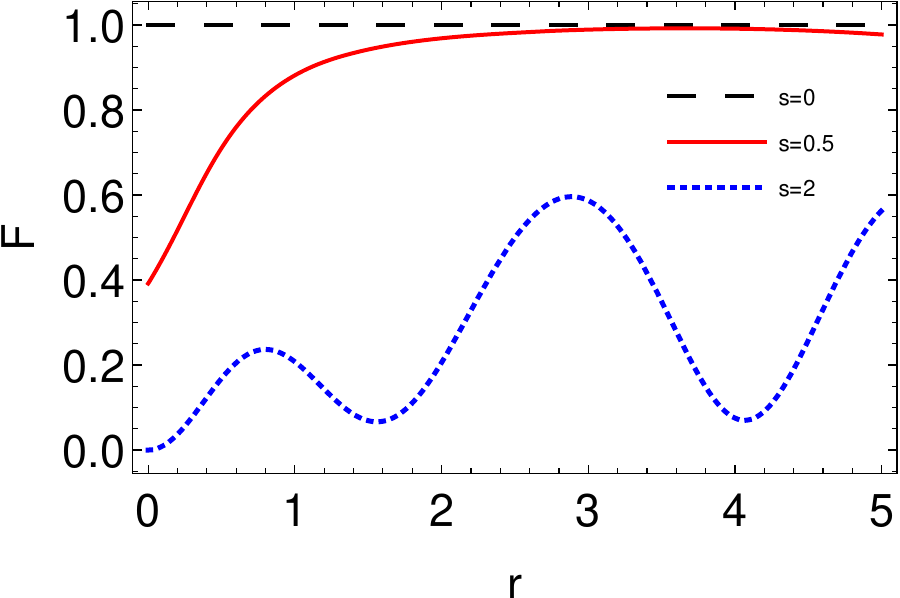}

\caption{\label{Fig:3}(Color online) The state distance between $\vert\Phi\rangle$
and initial SPACS $\vert\phi\rangle$ as a function of coherent state
parameter $r$ for various coupling strengths. Here, we set values
$\theta=\frac{\pi}{4}$, $\delta=\frac{\pi}{6}$, $\varphi=\frac{7\pi}{9}$. }

\end{figure}

The postselected measurement taken on polarization degree of freedom
of the beam could spoil the inherent properties presented in its spatial
part. Before we investigate the phase-space distribution of SPACS
after postselected von Neumanm measurement, we check the similarity
between the initial SPACS $\vert\phi\rangle$ and the state $\vert\Phi\rangle$
after measurement. The state distance between those two states can
be evaluated by
\begin{equation}
F=\vert\langle\phi\vert\Phi\rangle\vert^{2},\label{eq:23}
\end{equation}
and its value is bounded $0\le F\le1$. If $F=1$ ($F=0$), then the
two states are totally same (totally different). The $F$ in our case
can be calculated after substituting equations Eq. (\ref{eq:3-2})and
Eq. (\ref{eq:10-2}) into the Eq. (\ref{eq:23}), and the analytical
results are shown in Fig. \ref{Fig:3}. In Fig. \ref{Fig:3} we present
the state distance $F$ as a function of system parameter $r$ for
different coupling strengths with a fixed large weak value. As shown
in Fig. \ref{Fig:3}, in the weak coupling regime ($s=0.5$), the
state after the postselected measurement maintains similarity with
the the coherent state parameter $r$. However, with increasing the
measurement strength, the initial state $\vert\phi\rangle$ is spoiled
and the similarity between the pointer states before and after the
measurement is decreases. 

In order to further explain the squeezing effects of SPACS after postselected
von Neumann measurement, in the rest of this section we study the
Wigner function of $\vert\Phi\rangle$. The Wigner distribution function
is the closest quantum analogue of the classical distribution function
in phase space. According to the value of the Wigner function we can
intuitively determine the strength of its quantum nature, and the
negative value of the Wigner function proves the nonclassicality of
the quantum state. The Wigner function exists for any state, and it
is defined as the two-dimensional Fourier transform of the symmetric
order characteristic function. Thus, the Wigner function for the state
$\rho=\vert\Phi\rangle\langle\Phi\vert$ is written as \citep{Agarwal2013}
\begin{equation}
W(z)\equiv\frac{1}{\pi^{2}}\int_{-\infty}^{+\infty}\exp(\lambda^{\ast}z-\lambda z^{\ast})C_{W}(\lambda)d^{2}\lambda,\label{eq:35-1}
\end{equation}
where $C_{N}(\lambda)$ is the normal ordered characteristic function,and
is defined as 
\begin{equation}
C_{W}(\lambda)=Tr\left[\rho e^{\lambda a^{\dagger}-\lambda^{\ast}a}\right].\label{eq:34}
\end{equation}
Using the notation $\lambda^{\prime},$$\lambda^{\prime\prime}$ for
the real and imaginary parts of $\lambda$ and setting $z=x+ip$ to
emphasize the analogy between the radiation field quadratures and
the normalized dimensionless position and momentum observables of
the beam in phase space. We can rewrite the definition of the Wigner
function in terms of $x,p$ and $\lambda^{\prime},\lambda^{\prime\prime}$
as 
\begin{equation}
W(x,p)=\frac{1}{\pi^{2}}\int_{-\infty}^{+\infty}e^{2i(p\lambda^{\prime}-x\lambda^{\prime\prime})}C_{W}(\lambda)d\lambda^{\prime}d\lambda^{\prime\prime}.\label{eq:26}
\end{equation}
 By substituting the final normalized pointer state $\vert\Phi\rangle$
into Eq. (\ref{eq:26}), we can calculate the explicit expression
of its Wigner function and it reads as 
\begin{align}
W(z) & =\frac{2\vert\kappa\vert^{2}}{\pi(1+\vert\alpha\vert^{2})}e^{-2\vert z-\alpha\vert^{2}}\times\nonumber \\
 & \{\vert1+\langle\sigma_{x}\rangle_{w}\vert^{2}w(\Gamma)+\vert1-\langle\sigma_{x}\rangle_{w}\vert^{2}w(-\Gamma)\nonumber \\
 & \!\!\!\!\!\!\!\!\!+2\left(-1+\vert2z-\alpha\vert^{2}\right)Re[(1+\langle\sigma_{x}\rangle_{w})^{\ast}(1-\langle\sigma_{x}\rangle_{w})e^{2isIm[z]}]\}.\label{eq:35}
\end{align}
with 
\begin{align}
w(\Gamma) & =e^{-\frac{1}{2}s^{2}}e^{-2(Re[\alpha]-Re[z])s}\times\nonumber \\
 & \left(-1+\vert2z-\alpha\vert^{2}+2s(Re[\alpha]-2Re[z]+\frac{s}{2})\right)
\end{align}
 This is a real Wigner function and its value is bounded $-\frac{2}{\pi}\leq W(\alpha)\leq\frac{2}{\pi}$
in whole phase space.

\begin{widetext}

\begin{figure}
\includegraphics[width=6cm]{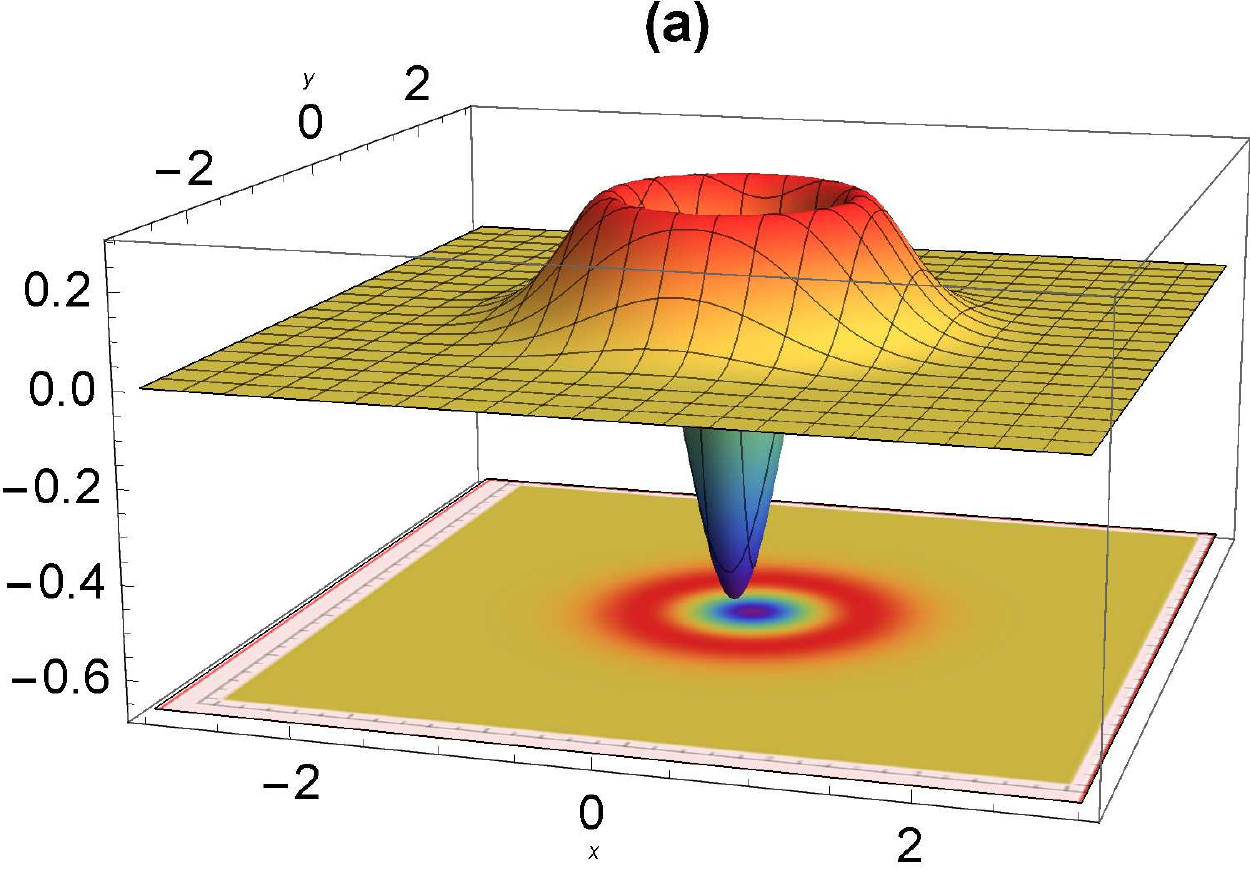}\includegraphics[width=6cm]{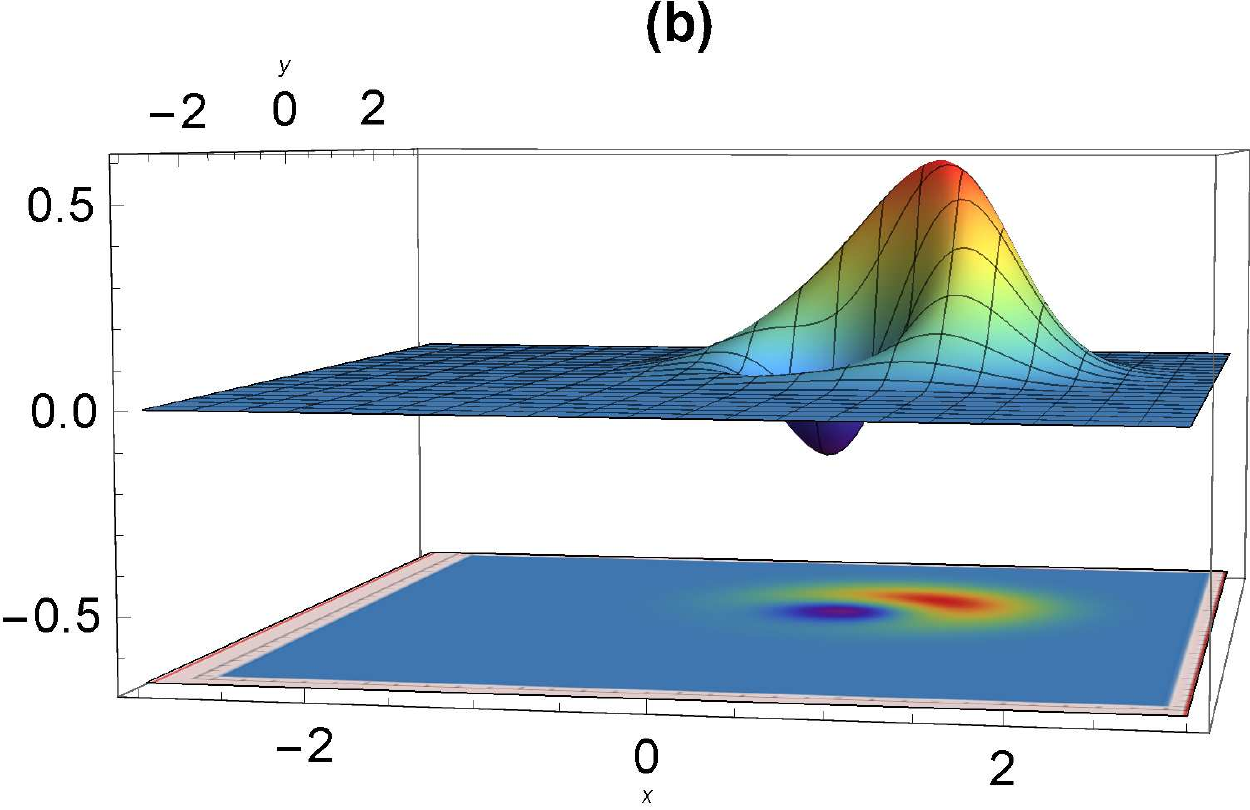}\includegraphics[width=6cm]{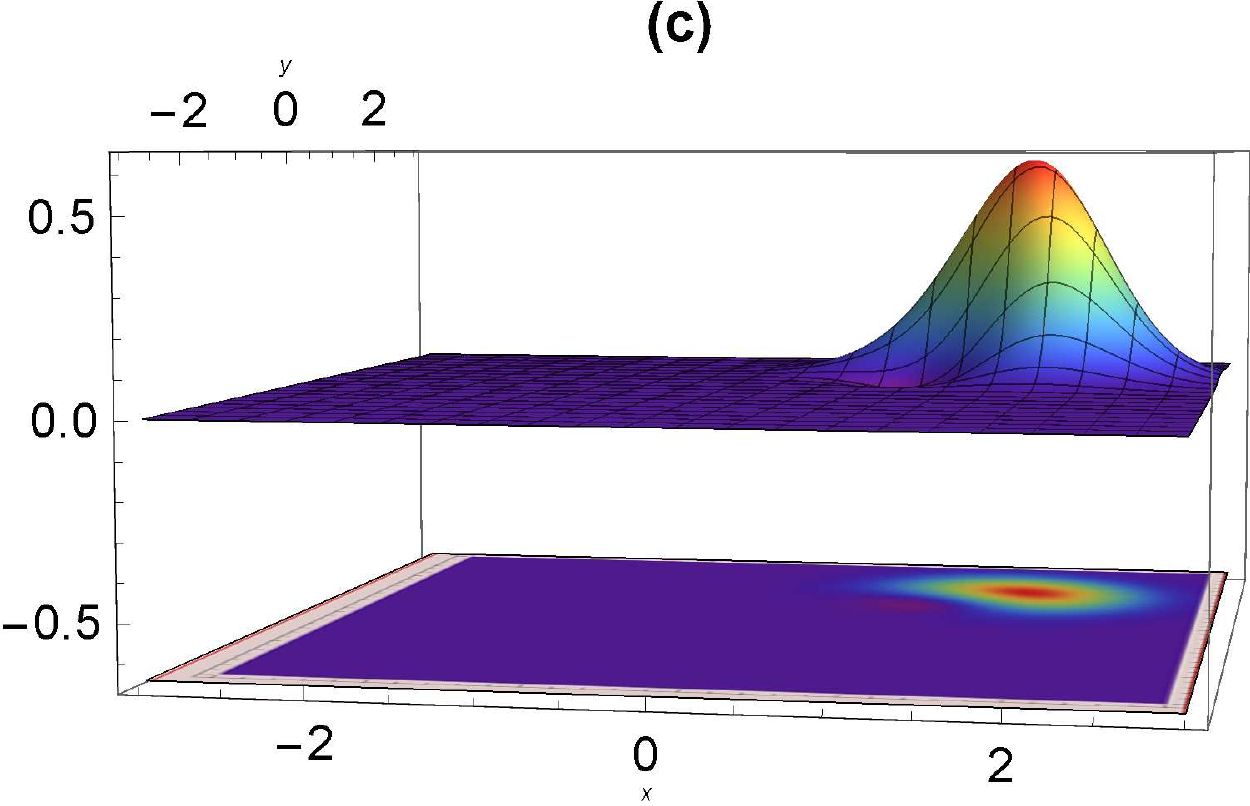}

\includegraphics[width=6cm]{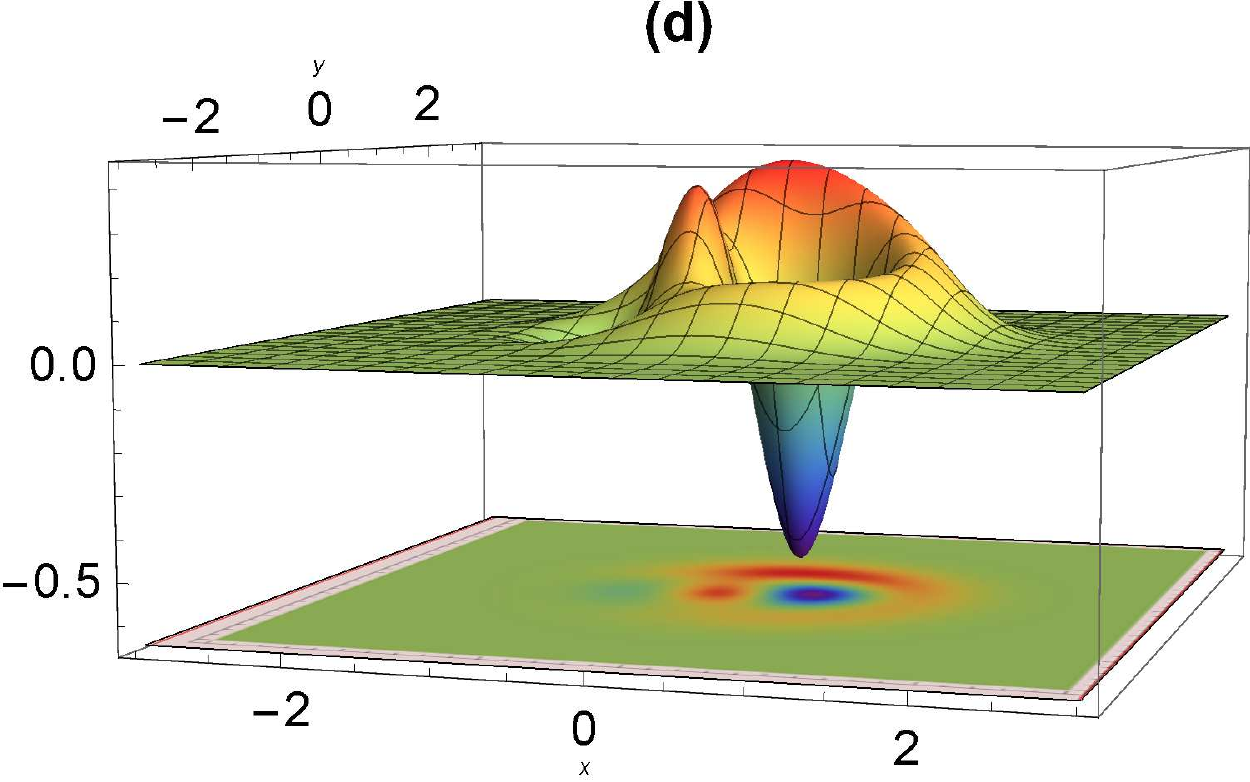}\includegraphics[width=6cm]{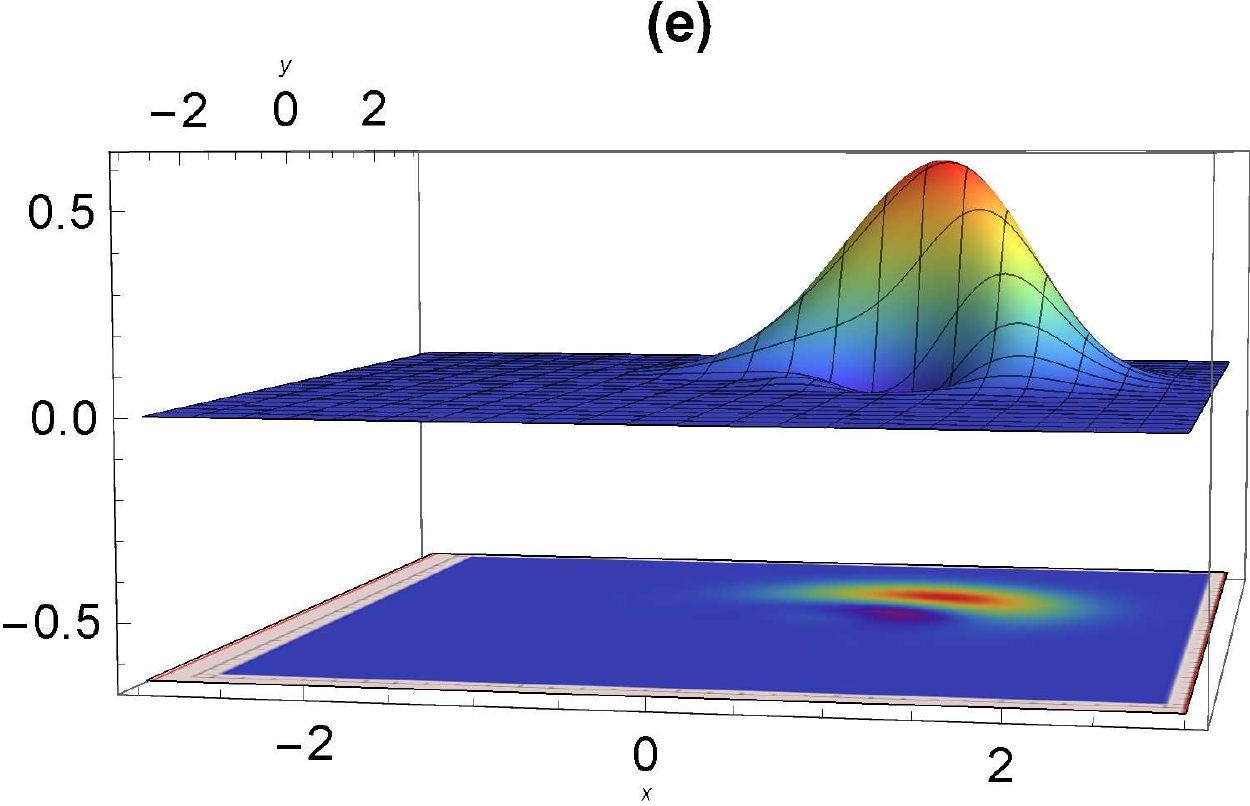}\includegraphics[width=6cm]{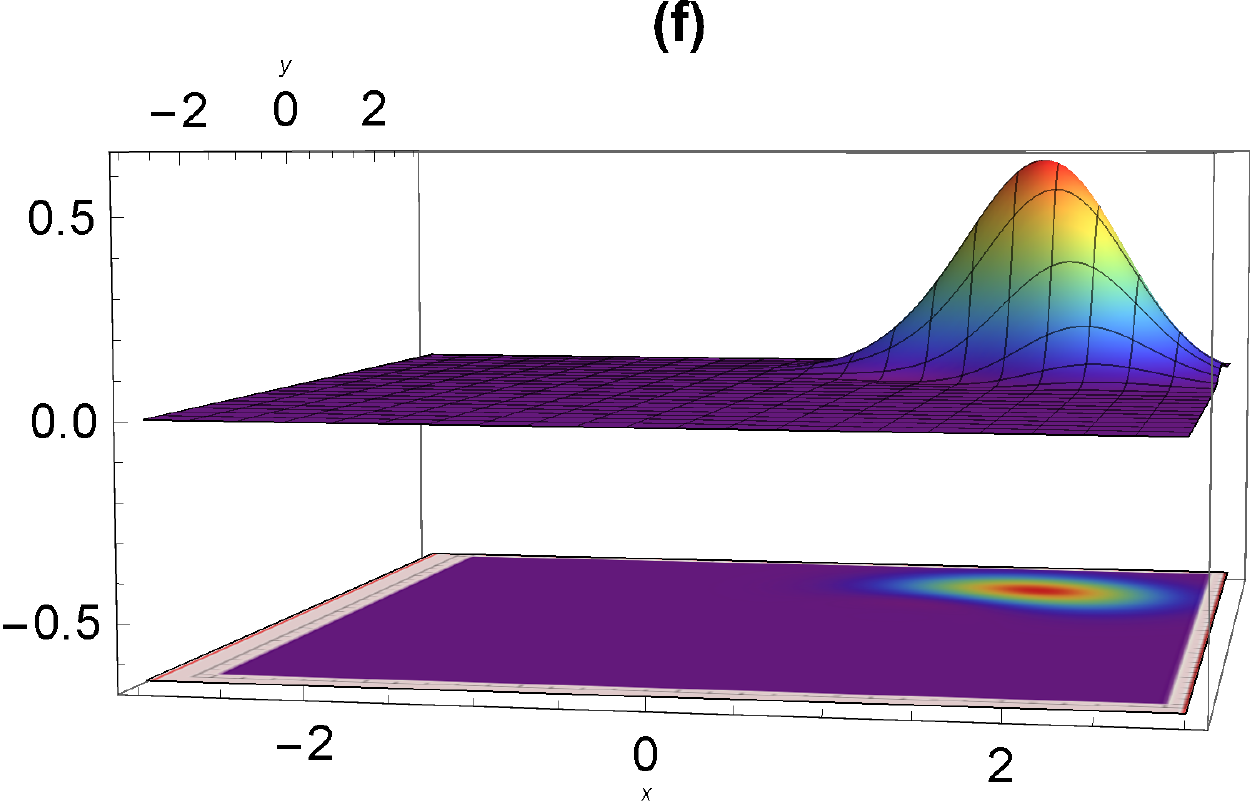}

\includegraphics[width=6cm]{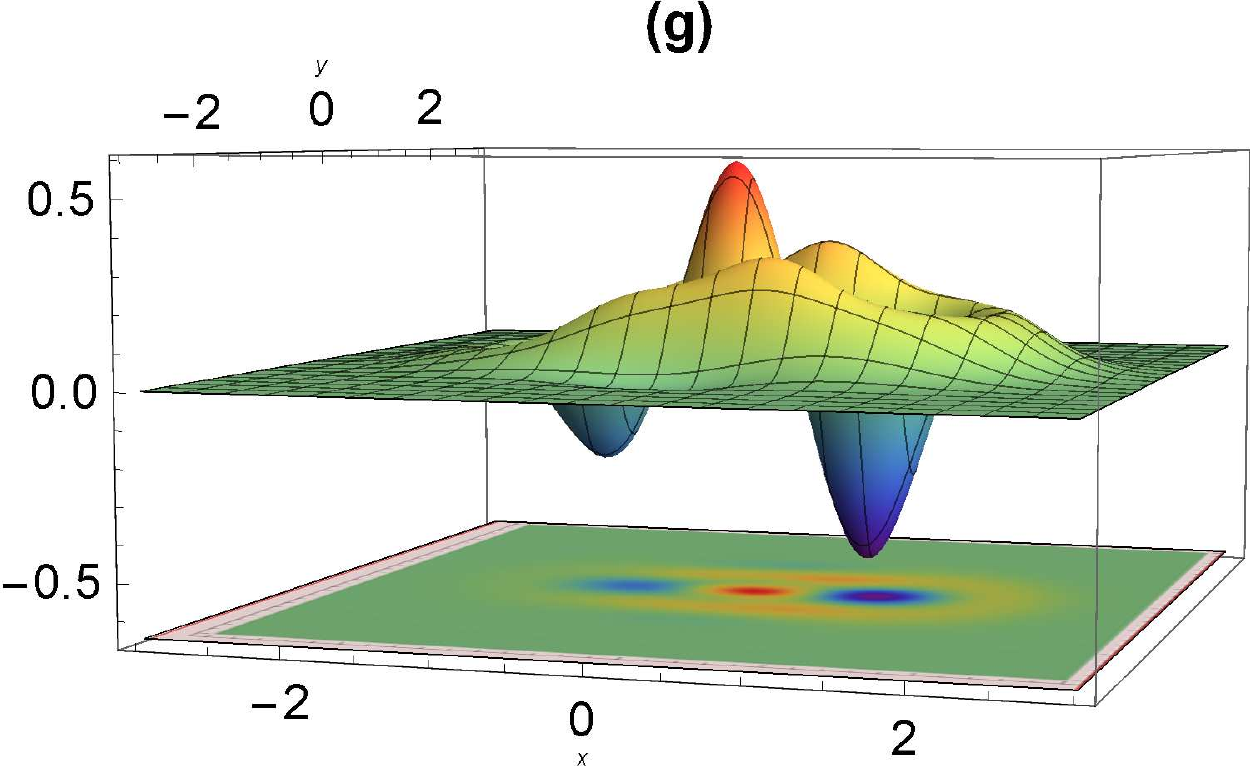}\includegraphics[width=6cm]{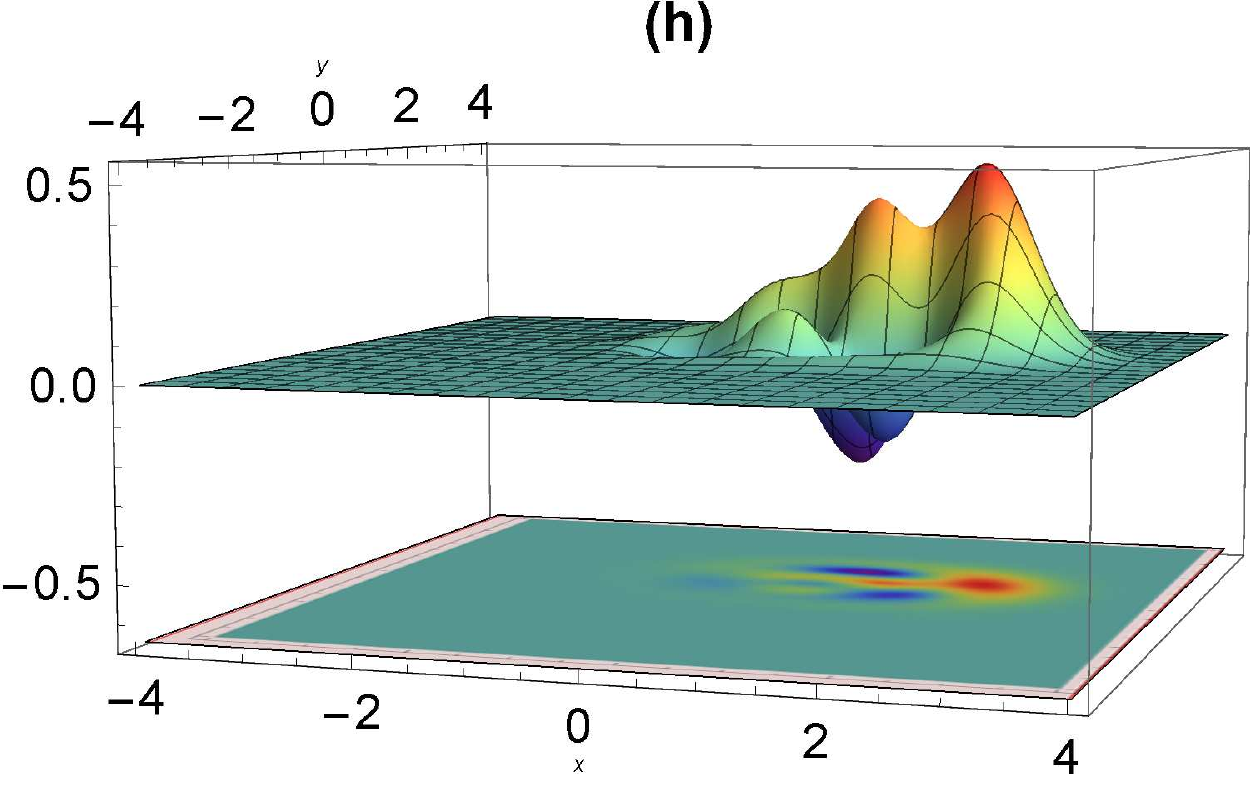}\includegraphics[width=6cm]{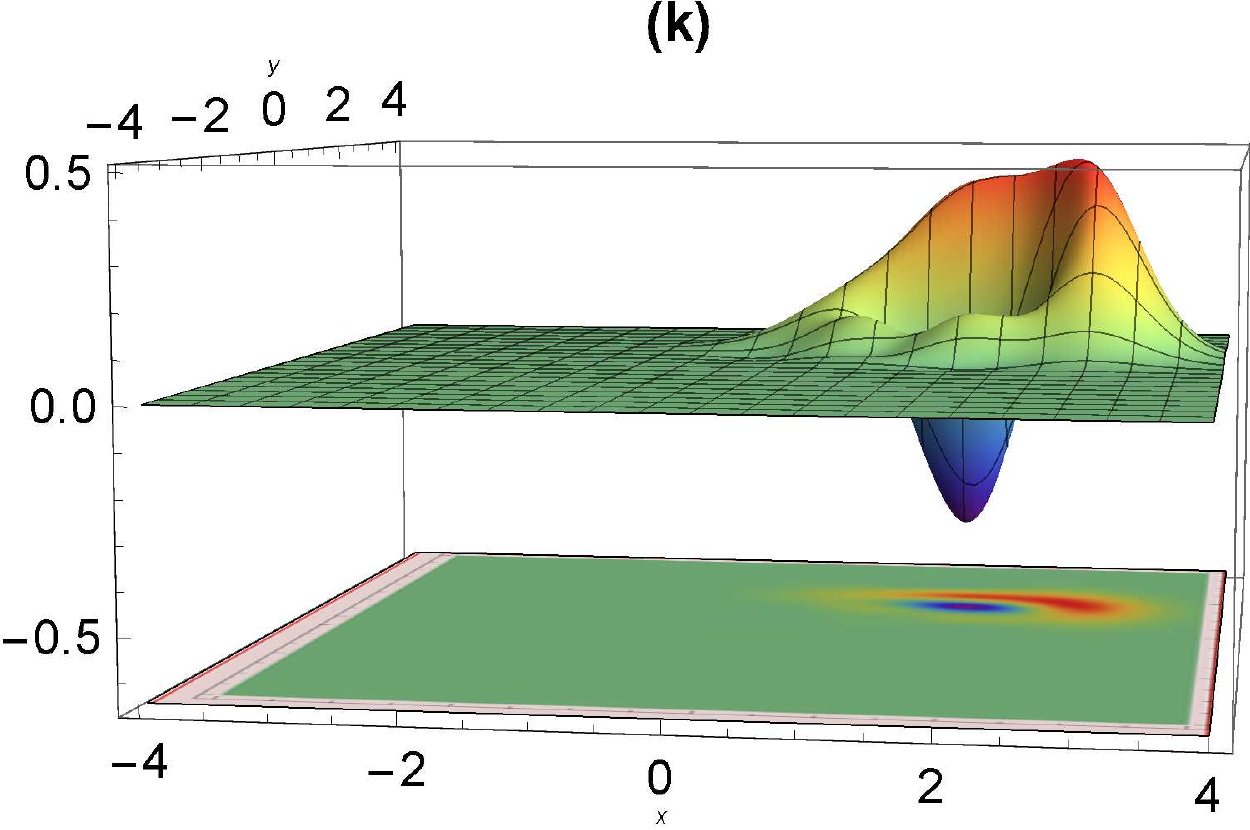}

\caption{\label{Fig:4}(Color online) Wigner function of SPACS with changing
parameters. Each column is defined for the different coherent state
parameter $\alpha$ with $r=0,1,2$, and are ordered accordingly from
left to right. Figures (a) to (c) correspond to $s=0$, (d) to (f)
correspond to $s=0.5$, and (g) to (k) correspond to $s=2$. Other
parameters are the same as those used in Fig. (\ref{Fig:3}).}
\end{figure}

\end{widetext}

To depict the effects of the postselected von Neumann measurement
on the non-classical feature of SPACS, in Fig. \ref{Fig:4} we plot
its curves for different parametric coherent state parameters $r$
and coupling strengths $s$. Each column from left to right in-turn
indicate the different coherent state parameters $r$ for 0, 1 and
2, and each row from up to down represent the different coupling strengths
$s$ for $0,0.5$ and $2$. It is observed that the positive peak
of the Wigner function moves from the center to the edge position
in phase space and its shape gradually becomes irregular with changing
coupling strength $s$. From the first row (see Figs. \ref{Fig:4}a-c
) we can see that the original SPACS exhibit inherent features changing
from single photon state to coherent states with gradually increasing
coherent state parameter $r$. Figs. \ref{Fig:4}d-k indicate the
phase space density function $W\left(z\right)$ after postselected
von Neumann measurement. Fig. \ref{Fig:4} d-f represent the Wigner
function for fixed weak interaction strength $s=0.5.$ It can be observed
that the Wigner function distribution shows squeezing in phase space
compared to the original SPACS and this kind of squeezing is pronounced
with increasing coupling strength (see Figs. \ref{Fig:4}g-k ). Furthermore,
in Figs. \ref{Fig:4}(g-k) we can see that in the strong measurement
regime significant interference structures manifest and the negative
regions become larger than the initial pointer state. 

As mentioned above, the existence of and progressively stronger negative
regions of the Wigner function in phase space indicates the degree
of nonclassicality of the associated state. From the above analysis
we can conclude that after the postselected von Neumann measurement,
the phase space distribution of SPACS is not only squeezed but the
nonclassicality is also pronounced in the strong measurement regime. 

\section{\label{sec:5}Conclusion }

In this paper we have studied the squeezing and Wigner function of
SPACS after postselected von Neumann measurement. In order to achieve
our goal, we first determined the final state of the pointer state
along with the standard measurement process. We examined the ordinary
(first-order) and ASS effects after measurement, and found that in
the weak measurement region, the ordinary squeezing and ASS of the
light field increased significantly as the weak value increased. 

To further explain our work, we examined the similarity between the
initial SPACS and the state after measurement. We observed that under
weak coupling, the state after the postselected measurement maintains
similarity with the initial state. However, as the intensity of the
measurement increases, the similarity between them gradually decreased
and indicated that the measurement spoils the system state if the
measurement is strong. We also investigate the Wigner function of
the system after postselected measurement. It is observed that following
the postselected von Neumann measurement, the phase space distribution
of SPACS is not only squeezed, but also adevelops significant interference
structures in the strongly measured regime. It also possess pronounced
nonclassicality characterized with a large negative area in phase
space.

We anticipate that the theoretical scheme in this paper may provide
an effective method for solving practical problems in quantum information
processing associated with SPACS.
\begin{acknowledgments}
This work was supported by the Natural Science Foundation of Xinjiang
Uyghur Autonomous Region (Grant No. 2020D01A72), the National Natural
Science Foundation of China (Grant No. 11865017) and the Introduction
Program of High Level Talents of Xinjiang Ministry of Science.
\end{acknowledgments}

\bibliographystyle{apsrev4-1}
\addcontentsline{toc}{section}{\refname}\bibliography{ref}

\end{document}